\documentclass[letterpaper, 12pt]{article}
\usepackage[english]{babel}
\usepackage{amsfonts,amsmath,amssymb,amsthm,latexsym,amscd,mathrsfs}
\usepackage{ifthen,cite}
\usepackage[bookmarksnumbered=true]{hyperref}
\usepackage{times,fancyhdr}

\usepackage{amsmath}
\numberwithin{equation}{section}

\usepackage{graphicx}

\hypersetup{pdfpagetransition={Split}}

\newcommand{\evenhead}{Author \ name}
\newcommand{\oddhead}{Article \ name}
\newcommand{\theArticleName}{Article \ name}

\newcommand{\FirstPageHeading}[1]{\thispagestyle{empty}%
\noindent\raisebox{0pt}[0pt][0pt]{\makebox[\textwidth]{\protect\footnotesize \sf }}\par}

\newcommand{\ArticleName}[1]{\renewcommand{\theArticleName}{#1}\vspace{-2mm}\par\noindent {\LARGE\bf  #1\par}}
\newcommand{\Author}[1]{\vspace{5mm}\par\noindent {\Large  #1\par} \par\vspace{2mm}\par}
\newcommand{\Address}[1]{\vspace{2mm}\par\noindent {\it #1} \par}
\newcommand{\Email}[1]{\ifthenelse{\equal{#1}{}}{}{\par\noindent {\rm E-mail: }{\it  #1} \par}}
\newcommand{\URLaddress}[1]{\ifthenelse{\equal{#1}{}}{}{\par\noindent {\rm URL: }{\tt  #1} \par}}
\newcommand{\EmailD}[1]{\ifthenelse{\equal{#1}{}}{}{\par\noindent {$\phantom{\dag}$~\rm E-mail: }{\it  #1} \par}}
\newcommand{\URLaddressD}[1]{\ifthenelse{\equal{#1}{}}{}{\par\noindent {$\phantom{\dag}$~\rm URL: }{\tt  #1} \par}}

\newcommand{\Abstract}[1]{\vspace{6mm}\par\noindent\hspace*{10mm}
\parbox{140mm}{\small {\bf Abstract.} #1}\par}
\newcommand{\Keywords}[1]{\vspace{3mm}\par\noindent\hspace*{10mm}
\parbox{140mm}{\small {\bf Key words:} \rm #1}\par}
\newcommand{\Classification}[1]{\vspace{3mm}\par\noindent\hspace*{10mm}
\parbox{140mm}{\small {\it 2020 Mathematics Subject Classification:} \rm #1}\vspace{3mm}\par}
\newcommand{\ShortArticleName}[1]{\renewcommand{\oddhead}{#1}}
\newcommand{\AuthorNameForHeading}[1]{\renewcommand{\evenhead}{#1}}

\long\def\@makecaption#1#2{
  \sbox\@tempboxa{\small \textbf{#1.}\ \ #2}%
  \ifdim \wd\@tempboxa >\hsize
    {\small \textbf{#1.}\ \ #2}\par \else
    \global \@minipagefalse
    \hb@xt@\hsize{\hfil\box\@tempboxa\hfil}%
  \fi \vskip\belowcaptionskip}

\def\numberwithin#1#2{\@ifundefined{c@#1}{\@nocounterr{#1}}{%
  \@ifundefined{c@#2}{\@nocnterr{#2}}{%
  \@addtoreset{#1}{#2}%
  \toks@\@xp\@xp\@xp{\csname the#1\endcsname}%
  \@xp\xdef\csname the#1\endcsname
    {\@xp\@nx\csname the#2\endcsname.\the\toks@}}}}
\def\E^#1{{\buildrel #1 \over\vee}}
\newtheorem*{theorem*}{Theorem}
\newtheorem{theorem}{Theorem}
\newtheorem{lemma}{Lemma}

{\theoremstyle{definition} 
\newtheorem*{criterion*}{Criterion}

}

\usepackage{tikz}

\begin{document}

\FirstPageHeading{V.I. Gerasimenko}

\ShortArticleName{Kinetic equations of open systems}

\AuthorNameForHeading{V.I. Gerasimenko}

\ArticleName{\textcolor{blue!50!black}{Towards kinetic equations of open systems \\ of active soft matter}
}

\Author{V.I. Gerasimenko}

\Address{Institute of Mathematics of the NAS of Ukra\"{\i}ne,\\
    \hspace*{3mm}3, Tereshchenkivs'ka Str.,\\
    \hspace*{3mm}01601, Ky\"{\i}v-4, Ukra\"{\i}ne\\
    E-mail: \emph{gerasym@imath.kiev.ua}}

\bigskip

\Abstract{The chapter presents some new approaches to describing the collective behavior of
complex systems of mathematical biology based on the evolution equations of observables such
as open systems. This representation of kinetic evolution has the look, in fact, to be the
most direct and mathematically fully consistent formulation modeling the collective behavior
of biological systems, since the traditionally used concept of the state in kinetic theory is
more subtle and is an implicit characteristic of the populations of living creatures. One of the
advantages of the developed approach is the opportunity to construct kinetic equations for open
complex systems in scaling approximations, involving initial correlations in particular, that
can characterize the condensed states of active soft matter. An approach is also related to
the challenge of a rigorous derivation of the non-Markovian kinetic equations from underlying
many-entity dynamics, which makes it possible to describe the memory effects of the collective
behavior of living creatures.
}

\bigskip

\Classification{35Q20; 47J35; 35Q92.}

\makeatletter
\renewcommand{\@evenhead}{
\hspace*{-3pt}\raisebox{-7pt}[\headheight][0pt]{\vbox{\hbox to \textwidth {\thepage \hfil \evenhead}\vskip4pt \hrule}}}
\renewcommand{\@oddhead}{
\hspace*{-3pt}\raisebox{-7pt}[\headheight][0pt]{\vbox{\hbox to \textwidth {\oddhead \hfil \thepage}\vskip4pt\hrule}}}
\renewcommand{\@evenfoot}{}
\renewcommand{\@oddfoot}{}
\makeatother

\newpage
\vphantom{math}

\protect\textcolor{blue!50!black}{\tableofcontents}

\vspace{0.8cm}

\textcolor{blue!50!black}{\section{Introduction: on evolution equations of active soft matter}}
Among the numerous complex systems studied by modern mathematical physics, significant progress has been
made over the past few decades in describing the collective behavior of systems with a large number of
living creatures (entities). Such systems include systems of large numbers of cells, colonies of bacteria,
populations of plants or swarms of animals \cite{ALBA,LP,SB-Y,TS18} and see references cited therein, as
well as similar examples in technology, such as robotics \cite{DK}. The approach to describing the evolution
of such systems at the moment is mainly associated with the continuous medium evolution equations of active
soft condensed matter \cite{MJRLPRAS,S20,B22,Li23}. Since the mentioned systems are significantly non-equilibrium,
a more consistent approach to describing their evolution is based on kinetic equations, the scale approximations
of which are the equations of evolution of an active soft medium \cite{ALBA,LP,SB-Y},\cite{G18,GF15,BL14}.
The purpose of this chapter is to present some new approaches to describing the collective behavior of open
complex systems by kinetic equations.

In modern research, the main approach to the problem of the rigorous derivation of kinetic equations consists
in the construction of suitable scaling asymptotics of solutions of evolution equations that describe the
evolution of states of a many-particle system, in particular, of asymptotics of a perturbative solution of
the Cauchy problem of the BBGKY (Bogolyubov--Born--Green--Kirkwood--Yvon) hierarchy \cite{GG23,CGP97,SR12}
(see also some recent papers \cite{GG18,GG21,PS16,D19,Ann23,GG22}).

The chapter examines open many-entity systems, since biological systems are never completely isolated from the
environment. When a biological system interacts with its environment, on a microscopic scale, its state tends
to become entangled with the larger number of living creatures in the environment, and this entanglement affects
what we can observe when we measure the system. Thereby, a rigorous description of decoherence phenomena in open
complex systems is interesting not only for fundamental reasons but is also crucial for practical applications
\cite{BK,B7,RH,K2,BGS}.

On the macroscopic scale, open complex systems of active soft condensed matter exhibit collective behavior that
differs from the statistical behavior of systems of many classical or quantum particles. Interacting with the
environment, the active soft matter can form new structures. For example, flocking and swarming are two typical
features of the collective behavior of a larger number of living creatures. The foremost distinctive specifics
of such systems are due to the fact that their constituents are self-propelled entities, in contrast to ordinary
gases or liquids, the particles of which move due to inertia.

Further, in this chapter, we review a new approach to the description of the collective behavior of open complex
systems within the framework of the evolution of observables \cite{GG18}. The proposed framework for the
description of kinetic evolution presents, in fact, the natural, fully consistent mathematical formulation
simulating the kinetic evolution of biological systems since the notion of the state is an implicit characteristic
of open systems of living creatures.

Note that one of the advantages of such an approach is the possibility to derive kinetic equations, involving
initial correlations of states of an open system, in particular, that characterize condensed states of active
soft matter. It will also make it possible to describe the mentioned above decoherence process in open complex
systems within the framework of the kinetic equations with initial correlations \cite{M23,CNB,ARB,PHW,AMA,GG15,GG14}.
We emphasize that this approach is also related to the problem of rigorously deriving non-Markovian generalizations
of the Fokker--Planck equations from the underlying dynamics of open complex systems, which make it possible to
describe memory effects in the system of living creatures interacting with the environment \cite{M23}.

In the next section, we will establish the hierarchy of evolution equations for reduced observables of an open
system of interacting stochastic processes of collisional kinetic theory \cite{L11,GF,GF14,G17}, modeling on the
microscopic scale the evolution of open systems of active soft matter. Then we will construct a non-perturbative
solution of the Cauchy problem to this hierarchy of evolution equations, which will form the basis for the
description of the evolution of states for considered open systems. In the next sections, we establish that for
initial states specified by means of a reduced distribution function of a system and correlation functions of an
open system, the evolution of reduced observables of a system can also be described by the Fokker--Planck type
kinetic equation with initial correlations, and the evolution of reduced observables of a system and an
environment is equivalent to the sequence of explicitly defined correlation functionals with respect to a solution
of the stated kinetic equation, which describe the processes of the propagation of correlations in an open system
of interacting stochastic processes.

\textcolor{blue!50!black}{\section{Hierarchy of evolution equations for observables\\
of open systems of self-propelled entities}}

As mentioned above, the main feature of active condensed matter is due to the fact that their constituents
are self-propelled entities, in contrast to ordinary gases or liquids, the particles of which move by inertia.
Further, following the works \cite{L11,GF,GF14,G17}, we will examine systems of interacting stochastic processes
of the collisional kinetic theory, modeling on a microscopic scale the evolution of open systems of active
condensed matter.

We consider an open many-entity system composed of a tracer stochastic process and an environment that is a
system of non-fixed, i.e., an arbitrary number of identical stochastic processes in the space $\mathbb{R}^{3}$.
Every $i$th entity in an environment is characterized by variables:
$\textbf{u}_i=(j_i,u_i)\in\mathcal{J}\times\mathcal{U}$, where $j_i\in\mathcal{J}\equiv(1,\ldots,M)$
is a number of its species (subpopulations), and $u_i\in\mathcal{U}\subset\mathbb{R}^{d}$ is its microscopic
specification, and the traced entity is characterized by $\textbf{u}=(j,u)\in\mathcal{J}\times\mathcal{U}$,
respectively.

Let $C_\gamma$ be the space of sequences $b=(b_{1+0},b_{1+1},\ldots,b_{1+n},\ldots)$ of measurable bounded
functions $b_{1+n}=b_{1+n}(\textbf{u},\textbf{u}_1,\ldots,\textbf{u}_n)$ that are symmetric with respect
to permutations of the arguments $\textbf{u}_1,\ldots,\textbf{u}_n$ and equipped with the norm
\begin{eqnarray*}
    &&\|b\|_{C_\gamma}=\max_{n\geq0}\frac{\gamma^n}{n!}\|b_{1+n}\|_{C_{1+n}}=
      \max_{n\geq0}\frac{\gamma^n}{n!}\max_{j,j_1,\ldots,j_n}\max_{u,u_1,\ldots,u_n}
      \big|b_{1+n}(\textbf{u},\textbf{u}_1,\ldots,\textbf{u}_n)\big|.
\end{eqnarray*}
Then the stochastic dynamics of self-propelled entities of an open system is described by the semigroup
of operators of the Markov jump processes defined in the space $C_\gamma$
\begin{eqnarray}\label{semi}
   &&\mathbb{R}_{+}\ni t\mapsto \bigoplus^\infty_{n=0}\, e^{t\big(\Lambda_{\mathfrak{s}}(\diamond)+
     \Lambda_{\mathfrak{e}}(1,\ldots,n)+ \varepsilon\Lambda_{\mathfrak{int}}(\diamond,1,\ldots,n)\big)},
\end{eqnarray}
where the symbol $\diamond$ enumerates the entity of a system and the entities from the environment
are enumerated by $1,\ldots,n$. The generator $\Lambda_{\mathfrak{s}}(\diamond)+\Lambda_{\mathfrak{e}}(1,\ldots,n)+
\varepsilon\Lambda_{\mathfrak{int}}(\diamond,1,\ldots,n)$ of the operator semigroup (\ref{semi}) (the Liouville
operator of an open system of collisional kinetic theory \cite{L11}) is defined by the collision integrals of
the following structure: \\
the Liouville operator of a system
\begin{eqnarray}\label{gen_obs_s}
    &&\hskip-8mm\Lambda_{\mathfrak{s}}(\diamond)b_{1+n}(\textbf{u},\textbf{u}_1,\ldots,\textbf{u}_n)=
        \Lambda^{[1]}(\diamond)b_{1+n}(\textbf{u},\textbf{u}_1,\ldots,\textbf{u}_n)\doteq\\
    &&\hskip+4mm a^{[\diamond]}(\textbf{u})\big(\int\limits_{\mathcal{J}\times\mathcal{U}}
        A^{[1]}(\textbf{v};\textbf{u})b_{1+n}(\textbf{v},\textbf{u}_1,\ldots,\textbf{u}_n)d\textbf{v}-
        b_{1+n}(\textbf{u},\textbf{u}_1,\ldots,\textbf{u}_n)\big)\nonumber,
\end{eqnarray}
the Liouville operator of an environment
\begin{eqnarray}\label{gen_obs_en}
     &&\hskip-8mm\Lambda_{\mathfrak{e}}(1,\ldots,n)b_{1+n}(\textbf{u},\textbf{u}_1,\ldots,\textbf{u}_n)=\\
     &&\hskip+4mm (\sum_{i_1=1}^n\Lambda^{[1]}(i_1)+\sum_{m=2}^M \sum_{i_1\neq\ldots\neq i_m=1}^n
          \Lambda^{[m]}(i_1,\ldots,i_m))b_{1+n}(\textbf{u},\textbf{u}_1,\ldots,\textbf{u}_n)\doteq \nonumber\\
     &&\hskip+4mm \sum_{i_1=1}^n a^{[1]}(\textbf{u}_{i_1})\big(\int\limits_{\mathcal{J}\times\mathcal{U}}
        A^{[1]}(\textbf{v};\textbf{u}_{i_1})
        b_{1+n}(\textbf{u},\textbf{u}_1,\ldots,\textbf{u}_{i_1-1},\textbf{v},\textbf{u}_{i_1+1},
        \ldots,\textbf{u}_n)d\textbf{v}-\nonumber\\
     &&\hskip+4mm b_{1+n}(\textbf{u},\textbf{u}_1,\ldots,\textbf{u}_n)\big)+\nonumber\\
     &&\hskip+4mm \sum_{m=2}^M \sum_{i_1\neq\ldots\neq i_m=1}^n a^{[m]}(\textbf{u}_{i_1},\ldots,\textbf{u}_{i_m})
        \big(\int\limits_{\mathcal{J}\times\mathcal{U}}
        A^{[m]}(\textbf{v};\textbf{u}_{i_1},\ldots,\textbf{u}_{i_m})\times\nonumber\\
    &&\hskip+4mm b_{1+n}(\textbf{u},\textbf{u}_1,\ldots,\textbf{u}_{i_1-1},\textbf{v},\textbf{u}_{i_1+1},
        \ldots,\textbf{u}_n)d\textbf{v}-b_{1+n}(\textbf{u},\textbf{u}_1,\ldots,\textbf{u}_n)\big)\nonumber,
\end{eqnarray}
and the Liouville operator of the interaction of a system with their environment
\begin{eqnarray}\label{gen_obs_int}
     &&\hskip-8mm\Lambda_{\mathfrak{int}}(\diamond,1,\ldots,n)b_{1+n}(\textbf{u},\textbf{u}_1,\ldots,\textbf{u}_n)=\\
     &&\hskip+4mm \sum_{m=1}^{M-1}\sum_{i_1\neq\ldots\neq i_m=1}^n\Lambda^{[\diamond+m]}(\diamond,i_1,\ldots,i_m)
        b_{1+n}(\textbf{u},\textbf{u}_1,\ldots,\textbf{u}_n)\doteq\nonumber\\
     &&\hskip+4mm \sum_{m=1}^{M-1}\sum_{i_1\neq\ldots\neq i_m=1}^n a^{[\diamond+m]}(\textbf{u},\textbf{u}_{i_1},\ldots,\textbf{u}_{i_m})
        \big(\int\limits_{\mathcal{J}\times\mathcal{U}}
        A^{[\diamond+m]}(\textbf{v};\textbf{u},\textbf{u}_{i_1},\ldots,\textbf{u}_{i_m})\times\nonumber\\
     &&\hskip+4mm b_{1+n}(\textbf{v},\textbf{u}_1,\ldots,\textbf{u}_n)d\textbf{v}-
       b_{1+n}(\textbf{u},\textbf{u}_1,\ldots,\textbf{u}_n)\big),\nonumber
\end{eqnarray}
where $\varepsilon>0$ is a scaling parameter, the functions $a^{[m]}(\textbf{u}_{i_1},\ldots,\textbf{u}_{i_m}),\,m\geq1,$
characterize the interaction between entities. We assume that these functions are measurable positive bounded functions
on $(\mathcal{J}\times\mathcal{U})^n$ such that: $0\leq a^{[m]}(\textbf{u}_{i_1},\ldots,\textbf{u}_{i_m})\leq a^{[m]}_*,$
where $a^{[m]}_*$ is some constant. The functions $A^{[m]}(\textbf{v};$ $\textbf{u}_{i_1},\ldots,\textbf{u}_{i_m}),\,m\geq1$,
are measurable positive integrable functions which describe the probability of the transition of the $i_1$ entity in the
microscopic state $u_{i_1}$ to the state $v$ as a result of the interaction with entities in the states
$u_{i_2},\ldots,u_{i_m}$. For every $m\geq1$, the function $A^{[m]}(\textbf{v};\textbf{u}_{i_1},\ldots,\textbf{u}_{i_m})$
satisfy the condition:
$\int_{\mathcal{J}\times\mathcal{U}}A^{[m]}(\textbf{v};\textbf{u}_{i_1},\ldots,\textbf{u}_{i_m})d\textbf{v}=1$.
We refer to the article \cite{L11}, which gives examples of the functions $a^{[m]}$ and $A^{[m]}$ in relation
to specific active soft matter.

The following statement is true, describing the properties of mapping (\ref{semi}) in the space $C_\gamma$ of
sequences of measurable bounded functions \cite{GF15},\cite{GF}. \\

\begin{lemma}
One-parameter mapping (\ref{semi}) is a $\ast$-weak continuous semigroup of bounded operators in the space $C_\gamma$.
\end{lemma}

We note that in the case of $m=1$ generator (\ref{gen_obs_en}) has the form $\sum_{i_1=1}^n\Lambda^{[1]}_n(i_1)$, and
it describes the free stochastic evolution of entities in an environment. The case of $m\geq2$ corresponds to a system
with the $m$-body interaction of entities in the sense accepted in kinetic theory. Further, for simplicity in notation
and statements, we restrict ourselves mainly to the case of a two-body interaction, i.e., for two subspecies of entities,
$M=2$, but the results are extended to an arbitrary number of subspecies. In this case, the generator of mapping
(\ref{semi}) has the following structure:
\begin{eqnarray}\label{gen_2}
    &&\Lambda_{\mathfrak{s}}(\diamond)=\Lambda^{[1]}(\diamond),\\
    &&\Lambda_{\mathfrak{e}}(1,\ldots,n)=\sum_{i_1=1}^n\Lambda^{[1]}(i_1)+
       \sum_{i_1\neq i_2=1}^n\Lambda^{[2]}(i_1,i_2), \nonumber\\
    &&\Lambda_{\mathfrak{int}}(\diamond,1,\ldots,n)=\sum_{i_1=1}^n\Lambda^{[\diamond+1]}(\diamond,i_1).\nonumber
\end{eqnarray}

Open complex systems of many entities are described in terms of two notions: observables and a state. Accordingly,
there are two equivalent approaches for describing the evolution of an open system, namely by means of evolution
equations for the observables or for a state \cite{GG21}. Let us consider the challenge of describing kinetic
evolution within the framework of the evolution of observables of many colliding self-propelled entities that
model interacting living beings in the environment.

The observables of an open system with a non-fixed number of environment entities are the sequences
$O=(O_{1+0},O_{1+1},\ldots,O_{1+n},\ldots)$ of functions $O_{1+n}(\textbf{u},\textbf{u}_1,\ldots,\textbf{u}_n)$
defined on $(\mathcal{J}\times\mathcal{U})^{1+n}$. The evolution of observables is described by the sequence
$O(t)=(O_{1+0}(t,\textbf{u}),\ldots,O_{1+n}(t,\textbf{u},\textbf{u}_1,\ldots,\textbf{u}_n),\ldots)$
of the functions
\begin{eqnarray*}
   &&O_{1+n}(t)=e^{t\big(\Lambda_{\mathfrak{s}}(\diamond)+\Lambda_{\mathfrak{e}}(1,\ldots,n)+
        \varepsilon\Lambda_{\mathfrak{int}}(\diamond,1,\ldots,n)\big)}O_{1+n}^0, \quad n\geq0,
\end{eqnarray*}
that is a global-in-time solution to the Cauchy problem for a sequence of the Liouville equations,
or also known as the forward Kolmogorov equations:
\begin{eqnarray}\label{LK}
   &&\frac{\partial}{\partial t}O_{1+n}(t)=\big(\Lambda_{\mathfrak{s}}(\diamond)+
        \Lambda_{\mathfrak{e}}(1,\ldots,n)+
        \varepsilon\Lambda_{\mathfrak{int}}(\diamond,1,\ldots,n)\big)O_{1+n}(t),\\ \nonumber\\
   \label{LKi}
   &&O_{1+n}(t)\mid_{t=0}=O_{1+n}^0, \quad n\geq0.
\end{eqnarray}

The mean values (expectation values) of observables are determined by the following positive continuous
linear functional defined on the space $C_\gamma$:
\begin{eqnarray}\label{averageD}
     &&\hskip-8mm\langle O\rangle(t)=(I,D(0))^{-1}(O(t),D(0))\doteq\\
     &&\hskip+4mm (I,D(0))^{-1}\sum\limits_{n=0}^{\infty}\frac{1}{n!}
        \int\limits_{(\mathcal{J}\times\mathcal{U})^{1+n}}d\textbf{u}d\textbf{u}_1\ldots d\textbf{u}_{n}\,
        O_{1+n}(t)\,D_{1+n}^0,\nonumber
\end{eqnarray}
where $D(0)=(D_{1+0}^0,D_{1+1}^0,\ldots,D_{1+n}^0,\ldots)$ is a sequence of nonnegative functions $D_{1+n}^0$
defined on $(\mathcal{J}\times\mathcal{U})^{1+n}$ that describes a state of an open system with a non-fixed
number of entities in an environment at an initial instant and
$(I,D(0))=\sum_{n=0}^{\infty}\frac{1}{n!}\int_{(\mathcal{J}\times\mathcal{U})^{1+n}}d\textbf{u}d\textbf{u}_1\ldots
d\textbf{u}_{n}\,D_{1+n}^0$ is a normalizing factor (the grand canonical partition function \cite{CGP97}).

Let $L^{1}_{\alpha}=\oplus^{\infty}_{n=0}\alpha^n L^{1}_{1+n}$ be the space of sequences
$f=(f_{1+0},\ldots,f_{1+n},\ldots)$ of the summable functions $f_{1+n}(\textbf{u},\textbf{u}_1,\ldots,\textbf{u}_n)$
defined on $(\mathcal{J}\times\mathcal{U})^{1+n}$, that are symmetric with respect to permutations
of the arguments $\textbf{u}_1,\ldots,\textbf{u}_n$, and equipped with the norm:
\begin{eqnarray*}
   &&\hskip-8mm\|f\|_{L^{1}_\alpha}=\sum\limits_{n=0}^\infty\alpha^n\|f_{1+n}\|_{L^{1}_{1+n}}=
     \sum\limits_{n=0}^\infty\alpha^n\int\limits_{(\mathcal{J}\times\mathcal{U})^{1+n}}
     d\textbf{u}d\textbf{u}_1\ldots d\textbf{u}_{n}\,
     \big|f_{1+n}(\textbf{u},\textbf{u}_1,\ldots,\textbf{u}_n)\big|,
\end{eqnarray*}
where $\alpha>1$ is a parameter. Then for $D(0)\in L^{1}_{\alpha}$ and $O(t)\in C_\gamma$ mean value
functional (\ref{averageD}) exists and it determines a duality between observables and a state.

Since in open systems the environment consists of an infinite number of constituents, for a successive
description of its evolution, another approach is used for the description of observables and a state,
namely by the sequences of so-called reduced observables and reduced distribution functions. This approach
is based on an equivalent representation of the functional for the average values of observables (a mean
value functional) \cite{GG21}.

The possibility of redefining of  the mean value functional is a result of dividing the series of expression
(\ref{averageD}) by the series of the normalization factor. As a result, the mean value functional is represented
by the following series expansion:
\begin{eqnarray}\label{avmar}
     &&\hskip-8mm(I,D(0))^{-1}(O(t),D(0))=(B(t),F(0))\doteq\sum\limits_{s=0}^{\infty}\frac{1}{s!}
       \int\limits_{(\mathcal{J}\times\mathcal{U})^{1+s}}d\textbf{u}d\textbf{u}_1\ldots d\textbf{u}_{s}\,
       B_{1+s}(t)\,F_{1+s}^0,
\end{eqnarray}
where the sequence of reduced observables $B(t)=(B_{1+0}(t,\textbf{u}),B_{1+1}(t,\textbf{u},\textbf{u}_1),\ldots,
B_{1+s}(t,\textbf{u},\textbf{u}_1,\ldots,$ $\textbf{u}_s),\ldots)$ is determined
by the expansions:
\begin{eqnarray}\label{mo}
      &&\hskip-8mm B_{1+s}(t,\textbf{u},\textbf{u}_1,\ldots,\textbf{u}_s)\doteq
         \sum_{n=0}^s\,\frac{(-1)^n}{n!}\sum_{j_1\neq\ldots\neq j_{n}=1}^s
            O_{1+s-n}\big(t,\textbf{u},(\textbf{u}_1,\ldots,\textbf{u}_s)\setminus(\textbf{u}_{j_1},\ldots,
            \textbf{u}_{j_{n}})\big),\\
      &&\hskip-8mm  s\geq 0,\nonumber
\end{eqnarray}
and, respectively, the sequence $F(0)$ of reduced distribution functions is determined by the following
series expansion:
\begin{eqnarray*}
      &&\hskip-8mm F_{1+s}^{0}(\textbf{u},\textbf{u}_1,\ldots,\textbf{u}_s)\doteq
         (I,D(0))^{-1}\sum\limits_{n=0}^{\infty}\frac{1}{n!}\,
          \int\limits_{(\mathcal{J}\times\mathcal{U})^{n}} d\textbf{u}_{s+1}\ldots d\textbf{u}_{s+n}\,
          D_{1+s+n}^0(\textbf{u},\textbf{u}_1,\ldots,\textbf{u}_{s+n}),\\
      &&\hskip-8mm s\geq 0.
\end{eqnarray*}
If $F(0)\in L^{1}_{\alpha}$ and $B(0)\in C_\gamma$, then at $t\in \mathbb{R}$ the functional $(B(t),F(0))$
exists under the condition that: $\gamma>\alpha^{-1}$.

Note that, heuristically formulated at the moment \cite{MJRLPRAS}, the continuous medium equations for
active soft matter represent the evolution equations for the macroscopic scale limit of mean value
functionals (\ref{avmar}) for the specific reduced observables of open systems.

The evolution of an open system with a non-fixed number of environment entities within the framework
of reduced observables (\ref{mo}) is described by the Cauchy problem for the following recurrence
evolution equations (we will also refer to them as the dual BBGKY hierarchy \cite{G18},\cite{GG18}):
\begin{eqnarray}\label{dh}
   &&\hskip-12mm \frac{\partial}{\partial t}B_{1+s}(t,\textbf{u},\textbf{u}_1,\ldots,\textbf{u}_s)=\\
   &&\big(\Lambda_{\mathfrak{s}}(\diamond)+\Lambda_{\mathfrak{e}}(1,\ldots,s)+
        \varepsilon\Lambda_{\mathfrak{int}}(\diamond,1,\ldots,s)\big)
        B_{1+s}(t,\textbf{u},\textbf{u}_1,\ldots,\textbf{u}_s)+\nonumber\\
   &&\varepsilon\sum_{j=1}^s\Lambda^{[2]}(\diamond,j)
        B_{1+s-1}(t,\textbf{u},(\textbf{u}_1,\ldots,\textbf{u}_s)\setminus\textbf{u}_{j})+\nonumber\\
   &&\sum_{j_1\neq j_{2}=1}^s\sum_{i\in(j_1,j_{2})}\Lambda^{[2]}(j_1,j_{2})
        B_{1+s-1}(t,\textbf{u},(\textbf{u}_1,\ldots,\textbf{u}_s)\setminus\textbf{u}_{i}),\nonumber\\
       \nonumber\\
   \label{dhi}
   &&\hskip-12mm B_{1+s}(t,\textbf{u},\textbf{u}_1,\ldots,\textbf{u}_s)|_{t=0}=
       B_{1+s}^{0}(\textbf{u},\textbf{u}_1,\ldots,\textbf{u}_s), \quad s\geq 0,
\end{eqnarray}
where the operators $\Lambda_{\mathfrak{s}}$, $\Lambda_{\mathfrak{e}}$ and $\Lambda_{\mathfrak{int}}$ are
defined by formulas (\ref{gen_2}), respectively, and the functions $B_{1+s}^{0},\,s\geq 0,$ are initial data.

The sequence $B(t)=(B_{1+0}(t,\textbf{u}),\ldots,B_{1+s}(t,\textbf{u},\textbf{u}_1,\ldots,\textbf{u}_s),\ldots)$
representing a solution to the Cauchy problem for recurrence evolution equations (\ref{dh}),(\ref{dhi}) is determined
by the following expansions \cite{GG23d}:
\begin{eqnarray}\label{sdh}
   &&\hskip-12mm B_{1+s}(t,\textbf{u},\textbf{u}_1,\ldots,\textbf{u}_s)=
       \sum_{n=0}^s\,\frac{1}{n!}\sum_{j_1\neq\ldots\neq j_{n}=1}^s
       \mathfrak{A}_{1+n}(t,\{\diamond,(1,\ldots,j_1-1,\\
   &&j_1+1,\ldots,j_n-1,j_n+1,\ldots,s)\},\,j_1,\ldots,j_{n})\,B_{1+s-n}^{0}(\textbf{u},\textbf{u}_1,\ldots,\nonumber\\
   &&\textbf{u}_{j_1-1},\textbf{u}_{j_1+1},\ldots,\textbf{u}_{j_n-1},\textbf{u}_{j_n+1},\ldots,\textbf{u}_s),
       \quad s\geq 0,\nonumber
\end{eqnarray}
where the generating operators of these expansions are the corresponding-order cumulants (semi-invariants)
of semigroups of operators (\ref{semi}). They are determined by the formulas \cite{GG21}:
\begin{eqnarray}\label{cumulantd}
   &&\hskip-12mm\mathfrak{A}_{1+n}(t,\{\diamond,Y\setminus X\},\,X)\doteq\\
   &&\hskip-5mm\sum\limits_{\mathrm{P}:\,(\{\diamond,Y\setminus X\},\,X)={\bigcup}_i X_i}
       (-1)^{\mathrm{|P|}-1}({\mathrm{|P|}-1})!\prod_{X_i\subset \mathrm{P}}
       (e^{t(\Lambda_{\mathfrak{s}}+\Lambda_{\mathfrak{e}}+
       \varepsilon\Lambda_{\mathfrak{int}})})(\theta(X_i)),\nonumber\quad n\geq 0,\nonumber
\end{eqnarray}
where such abridged notations are introduced: the sets of indexes are denoted by $(\diamond,Y)\equiv(\diamond,1,\ldots,s)$,
$X\equiv(j_1,\ldots,j_{n})$, the set, consisting of one element of indices
$(\diamond,Y\setminus X)=(\diamond,1,\ldots,j_1-1,j_1+1,\ldots,j_n-1,j_n+1,\ldots,s)$, is denoted by the symbol
$\{\diamond,Y\setminus X\}$, and the symbol $\sum_{\mathrm{P}:\{\diamond,Y\setminus X\},X)={\bigcup}_i X_i}$
means the sum over all possible partitions $\mathrm{P}$ of the set of the indexes
$(\{\diamond,Y\setminus X\},\,X)$ into $|\mathrm{P}|$ nonempty mutually disjoint subsets
$X_i\subset(\{\diamond,Y\setminus X\},X)$.
The declustering mapping $\theta(\cdot)$ is defined as follows: $\theta(\{\diamond,Y\setminus X\},X)=(\diamond,Y)$.

The simplest examples of expansions for reduced observables (\ref{sdh}) have the following form:
\begin{eqnarray*}
   &&\hskip-5mm B_{1+0}(t,\textbf{u})=\mathfrak{A}_{1}(t,\diamond)B_{1+0}^{0}(\textbf{u}),\\
   &&\hskip-5mm B_{1+1}(t,\textbf{u},\textbf{u}_1)=\mathfrak{A}_{1}(t,\{\diamond,1\})
      B_{1+1}^{0}(\textbf{u},\textbf{u}_1)+\mathfrak{A}_{2}(t,\diamond,\,1)B_{1+0}^{0}(\textbf{u}),
\end{eqnarray*}
and, for generating operators (\ref{cumulantd}) of these expansions, respectively:
\begin{eqnarray*}
   &&\hskip-5mm \mathfrak{A}_{1}(t,\{\diamond\})=e^{t\Lambda_{\mathfrak{s}}(\diamond)},\\
   &&\hskip-5mm \mathfrak{A}_{1}(t,\{\diamond,1\})=e^{t(\Lambda_{\mathfrak{s}}(\diamond)+
        \Lambda_{\mathfrak{e}}(1)+
        \varepsilon\Lambda_{\mathfrak{int}}(\diamond,1))},\\
   &&\hskip-5mm \mathfrak{A}_{2}(t,\diamond,1)=e^{t(\Lambda_{\mathfrak{s}}(\diamond)+
        \Lambda_{\mathfrak{e}}(1)+
        \varepsilon\Lambda_{\mathfrak{int}}(\diamond,1))}-
        e^{t\Lambda_{\mathfrak{s}}(\diamond)}e^{t\Lambda_{\mathfrak{e}}(1)}.
\end{eqnarray*}

We note that the corresponding-order cumulants (semi-invariants) of operator semigroups (\ref{cumulantd})
are solutions of the dual cluster expansions for these operator semigroups (\ref{semi}) and are represented
by the following recursive relations:
\begin{eqnarray}\label{cexd}
   &&\hskip-8mm (e^{t(\Lambda_{\mathfrak{s}}+\Lambda_{\mathfrak{e}}+
        \varepsilon\Lambda_{\mathfrak{int}})})(\diamond,(1,\ldots,s)\setminus(j_1,\ldots,j_{n}),j_1,\ldots,j_{n})=\\
   &&\hskip+4mm \sum\limits_{\mathrm{P}:\,(\{\diamond,(1,\ldots,s)\setminus(j_1,\ldots,j_{n})\},\,j_1,\ldots,j_{n})=
       \bigcup_i X_i}\,\prod\limits_{X_i\subset\mathrm{P}}\mathfrak{A}_{|X_i|}(t,X_i),\quad n\geq 0,\nonumber
\end{eqnarray}
where the abbreviations adopted above in (\ref{cumulantd}) are used.

In fact, the following criterion holds.

\begin{criterion*}
\emph{A solution of the Cauchy problem of the dual BBGKY hierarchy (\ref{dh}),(\ref{dhi}) is represented 
by expansions (\ref{sdh}) if and only if the generating operators of expansions (\ref{sdh}) are solutions 
of cluster expansions (\ref{cexd}) of operator semigroups (\ref{semi}) of the Liouville equations (\ref{LK})
(the forward Kolmogorov equations).}
\end{criterion*}

The necessity condition means that cluster expansions (\ref{cexd}) are take place for operator semigroups
(\ref{semi}). These recurrence relations are derived from definition (\ref{mo}) of reduced observables,
provided that they are represented as expansions (\ref{sdh}) for the solution of the Cauchy problem
of the dual BBGKY hierarchy (\ref{dh}),(\ref{dhi}).

The sufficient condition means that the infinitesimal generator of a semigroup of operators  (\ref{sdh})
coincides with the generator of the sequence of recurrence evolution equations (\ref{dh}).

We remark also that expansion (\ref{sdh}) can also be represented in the form of the perturbation
(iteration) series \cite{GG21} as a result of applying analogs of the Duhamel equation to cumulants
of semigroups of operators (\ref{cumulantd}).

Under the conditions formulated above on the generator of a semigroup of operators (\ref{semi}),
in the space $C_{\gamma}$, the following existence theorem is true.

\begin{theorem}
For initial data $B(0)=(B_{1+0}^{0},B_{1+1}^{0},\ldots,B_{1+s}^{0},\ldots)\in C_{\gamma}$ the
sequence $B(t)=(B_{1+0}(t),$ $B_{1+1}(t),$ $\ldots,B_{1+s}(t),\ldots)$ of reduced observables 
given by expansions (\ref{sdh}) is a global-in-time classical solution of the Cauchy problem 
to the dual BBGKY hierarchy (\ref{dh}),(\ref{dhi}) of an open system with a non-fixed number 
of environment entities.
\end{theorem}

For the following, we observe that the two-component sequence of reduced observables corresponds to observables
of a certain structure, namely, the reduced observable $B^{(1)}(0)=(0,O_{1+0}(\textbf{u}),O_{0+1}(\textbf{u}_1),$
$0,\ldots)$ corresponds to an additive-type observable \cite{GG21}. If we consider the additive-type observables
as initial data (\ref{dhi}), then the structure of the expansion of solution  (\ref{sdh}) is simplified and takes
the form
\begin{eqnarray}\label{af}
     &&\hskip-8mm B_{1+s}^{(1)}(t,\textbf{u},\textbf{u}_1,\ldots,\textbf{u}_s)=
                  \mathfrak{A}_{1+s}(t,\diamond,1,\ldots,s)O_{1+0}(\textbf{u})+\\
     &&\hskip+4mm \sum_{j=1}^s\mathfrak{A}_{s}(t,\{\diamond,j\},1,\ldots,j-1,j+1,\ldots,s) O_{0+1}(\textbf{u}_j),
                  \quad s\geq 0.\nonumber
\end{eqnarray}

We emphasize that the certain scaling asymptotics of the solution to the Cauchy problem for the dual BBGKY
hierarchy (\ref{dh}),(\ref{dhi}) describes the kinetic evolution of open complex systems within the framework
of the evolution of observables. In the following sections, we will not consider various scaling approximations
of this hierarchy of evolution equations, as this, is beyond the scope of this article. We will restrict
ourselves to one of the challenges of kinetic theory  \cite{CGP97},\cite{SR12}, which is the background of
describing the evolution within the framework of the notion of the state of a system interacting with the
environment. More precisely, we will further focus on the problem of the origin of describing the evolution
of the state of open systems by means of a kinetic equation of the Fokker--Planck type.

\textcolor{blue!50!black}{\section{Propagation of correlations in open systems\\ of self-propelled entities}}

We now consider the connection between the evolution of observables of an open system of active soft matter
and the kinetic evolution of its state, described in terms of a reduced distribution function  of a system.
In the case of the initial state of an open system, which is specified by the reduced distribution function
of a system and the correlation functions of a system and an environment, the dual picture of the evolution
to the picture described by means of observables governed by the dual BBGKY hierarchy for reduced observables
is the picture of the evolution of a state described in terms of the non-Markovian Fokker--Planck kinetic
equation and by a sequence of explicitly defined functionals of a solution of this kinetic equation that
describe the evolution of all possible correlations in an open system. It should be noted that many approaches
to deriving kinetic equations \cite{CNB,ARB,PHW} traditionally make the assumption of factorization (chaos
condition) \cite{BK},\cite{B7} of the initial state between the system and the environment.

Since condensed open systems are characterized by correlations between the state of a system and the environment,
we will consider initial states specified by a reduced distribution function of a system and the correlation
functions with the environment, namely
\begin{eqnarray}\label{ch}
   &&\hskip-8mm F^{(c)}(0)=(F_{1+0}^{0}(\textbf{u}),g_{1+1}(\textbf{u},\textbf{u}_1)
       F_{0+1}^{0}(\textbf{u}_1)F_{1+0}^{0}(\textbf{u}),\ldots,\\
   &&\hskip+4mm g_{1+n}(\textbf{u},\textbf{u}_1,\ldots,\textbf{u}_n)F_{0+n}^{0}(\textbf{u}_1,\ldots,\textbf{u}_n)
       F_{1+0}^{0}(\textbf{u}),\ldots),\nonumber
\end{eqnarray}
where function $F_{1+0}^{0}$ is the reduced distribution function of a system at the initial instant,
functions $\{F_{0+n}^{0}\}_{n\geq1}$, are initial reduced distribution functions of the environment,
and functions $\{g_{1+s}\}_{s\geq1}$, are initial correlation functions of the state of a system and
its environment. We note that the macroscopic characteristics of fluctuations of observables on a
microscopic scale are directly determined by the reduced correlation functions.

To describe the evolution of initial state (\ref{ch}), we introduce some preliminary notions, namely
the dual operator semigroup to the semigroup (\ref{semi}) in the sense of bilinear form (\ref{avmar}),
as well as the cumulants of such operator semigroups.

The dual stochastic dynamics of self-propelled entities of an open system is described by the semigroup
of operators of the Markov jump processes defined in the space $L^1_{\alpha}$
\begin{eqnarray}\label{dsemi}
   &&\mathbb{R}_{+}\ni t\mapsto \bigoplus^\infty_{n=0}\, e^{t\big(\Lambda_{\mathfrak{s}}^\ast(\diamond)+
        \Lambda_{\mathfrak{e}}^\ast(1,\ldots,n)+\varepsilon\Lambda_{\mathfrak{int}}^\ast(\diamond,1,\ldots,n)\big)},
\end{eqnarray}
whereas above, the symbol $\diamond$ enumerates the entity of a system, and the entities from the environment
are enumerated by $1,\ldots,n$. The generator of the semigroup of operators (\ref{dsemi}) (the dual Liouville
operator of an open system) has the structure
\begin{eqnarray}\label{dgen}
    &&\hskip-8mm \Lambda^{\ast}_{\mathfrak{s}}(\diamond)+\Lambda^{\ast}_{\mathfrak{e}}(1,\ldots,n)+
         \varepsilon\Lambda^{\ast}_{\mathfrak{int}}(\diamond,1,\ldots,n)=\\
     &&\hskip+4mm\Lambda^{\ast[1]}(\diamond)+\sum_{i_1=1}^n\Lambda^{\ast[1]}(i_1)+ \sum_{i_1\neq i_2=1}^n
          \Lambda^{\ast[2]}(i_1,i_2)+ \varepsilon\sum_{i_1=1}^n\Lambda^{\ast[\diamond+1]}(\diamond,i_1),\nonumber
\end{eqnarray}
where the dual operators to operators (\ref{gen_obs_s}),(\ref{gen_obs_en}),(\ref{gen_obs_int}) in the sense
of the bilinear form (\ref{avmar}) are defined accordingly as follows:
\begin{eqnarray*}
    &&\hskip-8mm \Lambda^{\ast[1]}(\diamond)f_{1+n}(\textbf{u},\textbf{u}_1,\ldots,\textbf{u}_n)\doteq\\
    &&\hskip+4mm\int\limits_{\mathcal{J}\times\mathcal{U}}a^{[\diamond]}(\textbf{v})
        A^{[\diamond]}(\textbf{u};\textbf{v})f_{1+n}(\textbf{v},\textbf{u}_1,\ldots,\textbf{u}_n)d\textbf{v}-
        a^{[\diamond]}(\textbf{u})f_{1+n}(\textbf{u},\textbf{u}_1,\ldots,\textbf{u}_n)\nonumber,
\end{eqnarray*}
\begin{eqnarray*}
    &&\hskip-8mm \Lambda^{\ast[1]}(i_1)f_{1+n}(\textbf{u},\textbf{u}_1,\ldots,\textbf{u}_n)\doteq\\
    &&\hskip+4mm\int\limits_{\mathcal{J}\times\mathcal{U}}a^{[1]}(\textbf{v})
        A^{[1]}(\textbf{u}_{i_1};\textbf{v})f_{1+n}(\textbf{u},\textbf{u}_1,\ldots,
        \textbf{u}_{i_1-1},\textbf{v},\textbf{u}_{i_1+1}\ldots,\textbf{u}_n)d\textbf{v}-\\
    &&\hskip+4mm a^{[1]}(\textbf{u}_{i_1})f_{1+n}(\textbf{u},\textbf{u}_1,\ldots,\textbf{u}_n),\nonumber,
\end{eqnarray*}
\begin{eqnarray*}
    &&\hskip-8mm \Lambda^{\ast[2]}(i_1,i_2)f_{1+n}(\textbf{u},\textbf{u}_1,\ldots,\textbf{u}_n)\doteq\\
    &&\hskip+4mm\int\limits_{\mathcal{J}\times\mathcal{U}}a^{[2]}(\textbf{v},\textbf{u}_{i_2})
        A^{[2]}(\textbf{u}_{i_1};\textbf{v},\textbf{u}_{i_2})
        f_{1+n}(\textbf{u},\textbf{u}_1,\ldots,\textbf{u}_{i_1-1},\textbf{v},\textbf{u}_{i_1+1},
        \ldots\textbf{u}_n)d\textbf{v}-\\
    &&\hskip+4mm a^{[2]}(\textbf{u}_{i_1},\textbf{u}_{i_2})
        f_{1+n}(\textbf{u},\textbf{u}_1,\ldots,\textbf{u}_n),\nonumber
\end{eqnarray*}
\begin{eqnarray*}
    &&\hskip-8mm \Lambda^{\ast[\diamond+1]}(\diamond,i_1))f_{1+n}(\textbf{u},\textbf{u}_1,\ldots,\textbf{u}_n)\doteq\\
    &&\hskip+4mm\int\limits_{\mathcal{J}\times\mathcal{U}}a^{[\diamond+1]}(\textbf{v},\textbf{u}_{i_1})
        A^{[\diamond+1]}(\textbf{u};\textbf{v},\textbf{u}_{i_1})
        f_{1+n}(\textbf{v},\textbf{u}_1,\ldots,\textbf{u}_n)d\textbf{v}-\\
    &&\hskip+4mm a^{[\diamond+1]}(\textbf{u},\textbf{u}_{i_1})
        f_{1+n}(\textbf{u},\textbf{u}_1,\ldots,\textbf{u}_n)\nonumber,
\end{eqnarray*}
and the functions $A^{[\diamond]}, A^{[1]}, A^{[2]}; a^{[\diamond]}, a^{[1]}, a^{[2]}$ are defined above.

The following statement is true, describing the properties of mapping (\ref{dsemi}) in the space
$L^{1}_{\alpha}=\oplus^{\infty}_{n=0}\alpha^n L^{1}_{1+n}$ of sequences of summable functions \cite{GF}. \\

\begin{lemma}
In the space $L^{1}_{1+n}$ the one-parameter mapping (\ref{dsemi}) is a bounded strong continuous
semigroup of operators.
\end{lemma}

We remark that the sequence of functions
\begin{eqnarray*}
   &&D_{1+n}(t)=e^{t\big(\Lambda_{\mathfrak{s}}^{\ast}(\diamond)+\Lambda_{\mathfrak{e}}^{\ast}(1,\ldots,n)+
        \varepsilon\Lambda_{\mathfrak{int}}^{\ast}(\diamond,1,\ldots,n)\big)}D_{1+n}^0, \quad n\geq0,
\end{eqnarray*}
is a global-in-time solution to the Cauchy problem for a sequence of the dual Liouville equations
or of the Kolmogorov backward equations with initial data $D_{1+n}^0,\,n\geq 0$, of open systems.

The $(1+n)$-order cumulant (semi-invariant) of operator semigroups (\ref{dsemi}) is determined
by the expansion \cite{GG21}:
\begin{eqnarray}\label{dkymyl}
   &&\hskip-8mm \mathfrak{A}_{1+n}^\ast(t,\{\diamond,Y\},X\setminus Y)=\\
   &&\hskip+4mm \sum\limits_{\mathrm{P}:\,(\{\diamond,Y\},\,X\setminus Y)
     ={\bigcup\limits}_i X_i}(-1)^{|\mathrm{P}|-1}(|\mathrm{P}|-1)!
     \prod_{X_i\subset \mathrm{P}}(e^{t(\Lambda_{\mathfrak{s}}^{\ast}+
     \Lambda_{\mathfrak{e}}^{\ast}+\varepsilon\Lambda_{\mathfrak{int}}^{\ast})})(\theta(X_i)),\nonumber
\end{eqnarray}
where such abridged notations are introduced: the sets of indexes are denoted by $\diamond,$ $Y\equiv(1,\ldots,s)$
and $X\equiv(1,\ldots,s+n)$, the set $\{\diamond,Y\}$ consists of one element $(\diamond,Y)=(\diamond,1,\ldots,s)$,
and the symbol $\sum_{\mathrm{P}:(\{\diamond,Y\},X\setminus Y)={\bigcup}_i X_i}$
means the sum over all possible partitions $\mathrm{P}$ of the set of the indexes $(\{\diamond,Y\},X\setminus Y)$
into $|\mathrm{P}|$ nonempty mutually disjoint subsets $X_i\subset(\{\diamond,Y\},X\setminus Y)$.
The declustering mapping $\theta(\cdot)$ is defined as above: $\theta(\{\diamond,Y\},X\setminus Y)=(\diamond,Y)$.

The simplest examples of expansions (\ref{dkymyl}) have the following form:
\begin{eqnarray*}
   &&\hskip-5mm \mathfrak{A}^\ast_{1}(t,\{\diamond\})=e^{t\Lambda^\ast_{\mathfrak{s}}(\diamond)},\\
   &&\hskip-5mm \mathfrak{A}^\ast_{1}(t,\{\diamond,1\})=e^{t(\Lambda^\ast_{\mathfrak{s}}(\diamond)+
        \Lambda^\ast_{\mathfrak{e}}(1)+\varepsilon\Lambda^\ast_{\mathfrak{int}}(\diamond,1))},\\
   &&\hskip-5mm \mathfrak{A}^\ast_{2}(t,\diamond,1)=e^{t(\Lambda^\ast_{\mathfrak{s}}(\diamond)+
         \Lambda^\ast_{\mathfrak{e}}(1)+\varepsilon\Lambda^\ast_{\mathfrak{int}}(\diamond,1))}-
         e^{t\Lambda^\ast_{\mathfrak{s}}(\diamond)}e^{t\Lambda^\ast_{\mathfrak{e}}(1)}.
\end{eqnarray*}

Now let us formulate the main statement of this section.

If the initial state is specified by sequence (\ref{ch}), then for the mean value functional
(\ref{avmar}) of reduced observables (\ref{sdh}) the following representation holds:
\begin{eqnarray}\label{w}
    &&\big(B(t),F^{(c)}(0)\big)=\big(B(0),F(t\mid F_{1}(t))\big),
\end{eqnarray}
where $F(t\mid F_{1}(t))=(F_{1+0}(t),F_{1+1}(t\mid F_{1}(t)),\ldots,F_{1+s}(t\mid F_{1}(t)),\ldots)$ is
a sequence of reduced functionals of the state with respect to a reduced distribution function of a system
\begin{eqnarray}\label{F_1(t)}
    &&\hskip-8mm F_{1+0}(t,\textbf{u})=\sum\limits_{n=0}^{\infty}\frac{1}{n!}
        \int\limits_{(\mathcal{J}\times\mathcal{U})^{1+n}}d\textbf{u}d\textbf{u}_{1}\ldots d\textbf{u}_{n}
        \mathfrak{A}_{1+n}^\ast(t,\diamond,1,\ldots,n)g_{1+n}(\textbf{u},\\
   &&\hskip+4mm\textbf{u}_1,\ldots,\textbf{u}_{n})F_{0+n}^{0}(\textbf{u}_1,\ldots,\textbf{u}_n)
        F_{1+0}^{0}(\textbf{u}),\nonumber
\end{eqnarray}
and the generating operator $\mathfrak{A}_{1+n}^\ast(t)$ of series expansions (\ref{F_1(t)}) is the
$(1+n)$-order cumulant (\ref{dkymyl}) of operator semigroups (\ref{dsemi}).

The reduced functionals of the state are defined by the following series expansions:
\begin{eqnarray}\label{f}
   &&\hskip-8mm F_{1+s}\big(t,\textbf{u},\textbf{u}_1,\ldots,\textbf{u}_s \mid F_{1+0}(t)\big)\doteq\\
   &&\hskip+4mm \sum_{n=0}^{\infty}\frac{1}{n!}\int\limits_{(\mathcal{J}\times\mathcal{U})^n}d\textbf{u}_{s+1}\ldots
      d\textbf{u}_{s+n}\,\mathfrak{V}_{1+n}(t,\{\diamond,Y\},X\setminus Y)F_{1+0}(t,\textbf{u}),
      \quad s\geq1,\nonumber
\end{eqnarray}
where the abridged notations are used: $Y\equiv(1,\ldots,s)$,\,\,$X\setminus Y\equiv(s+1,\ldots,s+n)$
and the generating operators of the corresponding order $\mathfrak{V}_{1+n}(t),\,n\geq0$, of these series
expansions are defined by the following expansions:
\begin{eqnarray}\label{skrrc}
   &&\hskip-8mm \mathfrak{V}_{1+n}\bigl(t,\{\diamond,Y\},X\setminus Y\bigr)\doteq\\
   &&\hskip-2mm n!\,\sum_{k=0}^{n}\,(-1)^k\,\sum_{n_1=1}^{n}\ldots
               \sum_{n_k=1}^{n-n_1-\ldots-n_{k-1}}\frac{1}{(n-n_1-\ldots-n_k)!}\,
               \widehat{\mathfrak{A}}_{1+n-n_1-\ldots-n_k}(t,\{\diamond,Y\},\nonumber\\
   &&\hskip-2mm s+1,\ldots, s+n-n_1-\ldots-n_k)\prod_{j=1}^k\,\sum\limits_{\mbox{\scriptsize$\begin{array}{c}
       \mathrm{D}_{j}:Z_j=\bigcup_{l_j}X_{l_j},\\
       |\mathrm{D}_{j}|\leq s+n-n_1-\dots-n_j\end{array}$}}\frac{1}{|\mathrm{D}_{j}|!}\times\nonumber\\
   &&\hskip-2mm
       \sum_{i_1\neq\ldots\neq i_{|\mathrm{D}_{j}|}=1}^{s+n-n_1-\ldots-n_j}\,
       \prod_{X_{l_j}\subset \mathrm{D}_{j}}\,\frac{1}{|X_{l_j}|!}\,
       \widehat{\mathfrak{A}}_{1+|X_{l_j}|}(t,i_{l_j},X_{l_j}).\nonumber
\end{eqnarray}
In expansion (\ref{skrrc}) the symbol $\sum_{\mathrm{D}_{j}:Z_j=\bigcup_{l_j} X_{l_j}}$ means the sum over
all possible dissections of the linearly ordered set $Z_j\equiv(s+n-n_1-\ldots-n_j+1,\ldots,s+n-n_1-\ldots-n_{j-1})$
on no more than $s+n-n_1-\ldots-n_j$ linearly ordered subsets, and the $(1+n)th$-order scattering cumulant we
denoted by the operator:
\begin{eqnarray}\label{scacu}
   &&\hskip-8mm\widehat{\mathfrak{A}}_{1+n}(t,\{\diamond,Y\},X\setminus Y)\doteq
      \mathfrak{A}_{1+n}^\ast(t,\{\diamond,Y\},X\setminus Y)g_{1+s+n}
      \mathfrak{A}_{1}^\ast(t,\diamond)^{-1}\prod_{i=1}^{s+n}\mathfrak{A}_{1}^\ast(t,i)^{-1},
\end{eqnarray}
where the operator $\mathfrak{A}_{1+n}^\ast(t)$ is the $(1+n)th$-order cumulant of the operator semigroups
(\ref{dsemi}).

We provide some examples of expressions for the generating operators of series (\ref{f}) for reduced
functionals of the state:
\begin{eqnarray*}\label{rrrls}
   &&\hskip-8mm\mathfrak{V}_{1}(t,\{\diamond,Y\})=\widehat{\mathfrak{A}}_{1}(t,\{\diamond,Y\})\doteq\\
   &&\hskip+4mme^{t(\Lambda_{\mathfrak{s}}^{\ast}(\diamond)+
       \Lambda_{\mathfrak{e}}^{\ast}(Y)+\varepsilon\Lambda_{\mathfrak{int}}^{\ast}(\diamond,Y))}
       g_{1+s}e^{-t\Lambda^{\ast[\diamond]}(\diamond)}\prod_{i=1}^{s}e^{-t\Lambda^{\ast[1]}(i)},\\
   &&\hskip-8mm\mathfrak{V}_{2}(t,\{\diamond,Y\},s+1)=\widehat{\mathfrak{A}}_{2}(t,\{\diamond,Y\},s+1)-
       \widehat{\mathfrak{A}}_{1}(t,\{\diamond,Y\})\sum_{i_1=1}^s\widehat{\mathfrak{A}}_{2}(t,i_1,s+1).
\end{eqnarray*}

If $\|F_{1+0}(t)\|_{L^{1}(\mathcal{J}\times\mathcal{U})}<e^{-(3s+2)}$, then for arbitrary $t\in \mathbb{R}$
series expansion (\ref{f}) converges in the norm of the space $L^{1}_{1+s}$.

Let us note that in the particular case of initial data (\ref{dhi}) specified by the additive-type
reduced observables, according to solution expansion (\ref{af}), equality (\ref{w}) takes the form
\begin{eqnarray*}\label{avmar-11}
   &&\hskip-8mm\big( B^{(1)}(t),F^{(c)}(0)\big)=
      \int\limits_{\mathcal{J}\times\mathcal{U}}d\textbf{u}\,O_{1+0}(\textbf{u})F_{1+0}(t,\textbf{u})+
      \int\limits_{(\mathcal{J}\times\mathcal{U})^2}d\textbf{u}d\textbf{u}_1\,O_{0+1}(\textbf{u}_1)
       F_{1+1}\big(t,\textbf{u},\textbf{u}_1\mid F_{1+0}(t)\big),\nonumber
\end{eqnarray*}
where the reduced distribution function $F_{1+0}(t)$ is determined by series (\ref{F_1(t)}). In the
case of initial data (\ref{dhi}) specified by the $s\geq2$-ary, reduced observable of the environment
equality (\ref{w}) has the form
\begin{eqnarray*}\label{avmar-12}
   &&\hskip-8mm\big(B^{(s)}(t),F^{(c)}(0)\big)=
      \frac{1}{s!}\,\int\limits_{(\mathcal{J}\times\mathcal{U})^s}d\textbf{u}
      d\textbf{u}_{1}\ldots d\textbf{u}_{s}\,O_{0+s}(\textbf{u}_1,\ldots,\textbf{u}_s)
      F_{1+s}(t,\textbf{u},\textbf{u}_1,\ldots,\textbf{u}_{s}\mid F_{1}(t)),\nonumber
\end{eqnarray*}
where the reduced functionals of the state $F_{1+s}(t\mid F_{1}(t)),\, s\geq1,$ are determined
by series expansion (\ref{f}).

The proof of representation (\ref{w}) of the mean value functional of reduced observables (\ref{sdh})
or their specific types is based on the application of cluster expansions to generating operators
(\ref{cumulantd}) of expansions (\ref{sdh}), which are dual to the kinetic cluster expansions introduced
in the paper \cite{GG}. Then the adjoint series expansion can be expressed through the reduced distribution
function (\ref{F_1(t)}) in the form of a functional from the right side of the equality (\ref{w}).

Thus, if the initial state is completely specified by a reduced distribution function and a sequence
of correlation functions (\ref{ch}), then, using a non-perturbative solution of the dual BBGKY hierarchy
(\ref{sdh}), was proven that all possible states at an arbitrary moment of time can be described within
the framework of the reduced distribution function (\ref{F_1(t)}), i.e., without any approximations.

We also emphasize that reduced functionals of the state (\ref{f}) in the case of $s\geq1$ characterize
the processes of the creation of correlations generated by the dynamics of open many-constituent systems
of active soft matter and the propagation of initial correlations.

\textcolor{blue!50!black}{\section{The Fokker--Planck generalized kinetic equation\\ with initial correlations}}
Now let us establish the evolution equation for the reduced distribution function (\ref{F_1(t)}),
i.e., the first element of the sequence $F(t\mid F_{1}(t))$, which describes the state of the
stochastic tracer process interacting with the environment.

As a result of differentiation with respect to the time variable of series expansion (\ref{F_1(t)})
in the sense of pointwise convergence in the space $L^{1}(\mathcal{ J}\times\mathcal{U})$, and then
of applying kinetic cluster expansions \cite{GG} for the cumulants of operator semigroups (\ref{dgen}),
the following identity is derived:
\begin{eqnarray}
  \label{ke}
    &&\hskip-8mm\frac{\partial}{\partial t}F_{1+0}(t,\textbf{u})=
       \Lambda^{\ast[\diamond]}(\diamond)F_{1+0}(t,\textbf{u})+
       \varepsilon \int\limits_{\mathcal{J}\times\mathcal{U}}d\textbf{u}_{1}
       \Lambda^{\ast[\diamond+1]}(\diamond,1)
       F_{1+1}\big(t,\textbf{u},\textbf{u}_{1}\mid F_{1+0}(t)\big),
\end{eqnarray}
where the functional $F_{1+1}(t\mid F_{1+0}(t))$ in the collision integral is determined by series
expansion (\ref{f}). The series expansion of the collision integral in equality (\ref{ke}) converges
in the sense of the norm convergence of the space $L^{1}(\mathcal{J}\times\mathcal{U})$ if for the
coefficients of the collision integral the following condition holds: $\alpha<e^{-4}$, where
$\sup_{n\geq0}\alpha^{-n}\|F^0_{0+n}\|_{L_{0+n}^1}<+\infty$.

We will treat the obtained identity (\ref{ke}) as the evolution equation for the reduced distribution
function of a tracer stochastic process interacting with an environment that is a system of non-fixed,
i.e., an arbitrary number of identical stochastic processes. We will refer to this evolution equation
as the Fokker--Planck generalized kinetic equation with initial correlations.

Taking into account the definition of generators (\ref{dgen}), such a kinetic equation is rewritten
in the following explicit form:
\begin{eqnarray}\label{gke}
    &&\hskip-8mm\frac{\partial}{\partial t}F_{1+0}(t,\textbf{u})=
       \int\limits_{\mathcal{J}\times\mathcal{U}}a^{[\diamond]}(\textbf{v})
        A^{[\diamond]}(\textbf{u};\textbf{v})F_{1+0}(t,\textbf{v})d\textbf{v}-
        a^{[\diamond]}(\textbf{u})F_{1+0}(t,\textbf{u})+\\
    &&\hskip+4mm\varepsilon \int\limits_{\mathcal{J}\times\mathcal{U}}d\textbf{u}_{1}\big(
       \int\limits_{\mathcal{J}\times\mathcal{U}}a^{[\diamond+1]}(\textbf{v},\textbf{u}_{1})
       A^{[\diamond+1]}(\textbf{u};\textbf{v},\textbf{u}_{1})
       F_{1+1}\big(t,\textbf{v},\textbf{u}_{1}\mid F_{1+0}(t)\big)d\textbf{v}-\nonumber\\
    &&\hskip+4mm a^{[\diamond+1]}(\textbf{u},\textbf{u}_{1})
       F_{1+1}\big(t,\textbf{u},\textbf{u}_{1}\mid F_{1+0}(t)\big)\big),\nonumber\\ \nonumber\\
  \label{gkei}
    &&\hskip-8mm F_{1+0}(t,\textbf{u})|_{t=0}=F_{1+0}^{0}(\textbf{u}).
\end{eqnarray}
where the functional $F_{1+1}(t\mid F_{1+0}(t))$ in the collision integral is determined by series
expansion (\ref{f}), and the functions $a^{[\diamond+1]}$ and $A^{[\diamond+1]}$ were defined above.

Under the conditions on the generator of operator semigroup formulated above (\ref{dsemi}), the
following statement is true. \\

\begin{theorem}
For small initial data $F_{1+0}^{0}\in L^{1}(\mathcal{J}\times\mathcal{U})$, series expansion
(\ref{F_1(t)}) is a global-in-time solution of the Cauchy problem of the Fokker--Planck generalized
kinetic equation with initial correlations (\ref{gke}),(\ref{gkei}).
For initial data $F_{1+0}^{0}\in L^{1}(\mathcal{J}\times\mathcal{U})$ it is a strong (classical)
solution, and for arbitrary initial data, it is a weak (generalized) solution.
\end{theorem}

We note that for initial states (\ref{ch}) specified by a reduced distribution function, the evolution
of the state of a system interacting with the environment described within the framework of a reduced
distribution function governed by the Fokker--Planck generalized kinetic equation with initial correlations
(\ref{gke}) is dual to the hierarchy of evolution equations (\ref{dh}) for reduced observables of a system
with respect to bilinear form (\ref{averageD}), and it is completely equivalent to the description of states
in terms of reduced distribution functions governed by the BBGKY hierarchy for open systems \cite{GG14}.

Thus, in the circumstances where the initial state of an open system is specified by the reduced
distribution function of a system and the correlation functions with an environment, for the mean
value functional of observables at an arbitrary instant, the representation is also valid in terms
of the reduced distribution function that describes the evolution of the state of an open system
governed by the non-Markovian kinetic equation.

In other words, it has been established that a generalization of the Fokker-Planck kinetic equation
provides an alternative approach to describing the evolution of the state of a self-propelled entity
in the environment. Note that specific kinetic equations of the Fokker--Planck type can be derived
based on the constructed generalization of the Fokker--Planck kinetic equation within appropriate
scaling limits or as a result of certain approximations.

\newpage

\textcolor{blue!50!black}{\section{Conclusion and outlook}}
At the outset, we note that complex systems as a discipline is an emerging field of science that studies
how the constituent systems and their interactions give rise to the collective behavior of the system
and how the system interacts with its environment. In this chapter, we introduced concepts that guide
understanding the origin of the description of complex systems by means of kinetic equations.

The chapter presents an approach to describing kinetic evolution within the framework of the evolution
of observable systems of many self-propelled entities that simulate interacting living creatures in
the environment. The proposed framework for describing evolution, in fact, is a direct, mathematically
completely consistent formulation of the kinetic evolution of open systems of living creatures.

One of the advantages of the suggested approach in comparison with the conventional approach of the
kinetic theory \cite{CGP97},\cite{SR12} consists in the possibility to construct the Fokker-Planck
type kinetic equations in various scaling limits in the presence of initial correlations, which
characterize the condensed states of interacting entities in open complex systems.

In the case of initial states (\ref{ch}) completely specified by reduced distribution functions
and correlation functions with the environment, based on the non-perturbative solution (\ref{sdh})
of the dual hierarchy of evolution equations (\ref{dh}) for reduced observables of an open system,
it is proved that all possible states of the system at an arbitrary moment of time are described
by a reduced distribution function governed by a non-Markovian generalization of the kinetic equation
Fokker--Planck with initial correlations (\ref{gke}), or, in other words, they can be described without
any approximations. Moreover, in the case of the initial states under consideration, the processes of
creating correlations and propagating initial correlations generated by the dynamics of open systems
of active soft matter are described by the functionals of the state (\ref{f}) constructed above, which
are governed by the derived kinetic equation with initial correlations.

Finally, we emphasize that beyond to the approach presented above, related to the problem of rigorous
derivation of generalizations of the Fokker--Planck equation from the underlying stochastic dynamics
of entities described by the semigroup of Markov jump processes  of open complex systems this approach
can also be applied to other types of dynamics of many self-propelled entities on a microscopic scale.

\vskip+7mm

\noindent \textbf{Acknowledgements.} \,\,{\large\textcolor{blue!55!black}{Glory to Ukra\"{\i}ne!}}

\bigskip

\addcontentsline{toc}{section}{\textcolor{blue!55!black}{References}}

\vskip+5mm

\vskip+8mm
\centerline{\textcolor{blue!55!black}{\textbf{************}}}

\end{document}